\documentclass[10pt,doublecolumn]{IEEEtran}

\pdfoutput=1
\usepackage[final]{pdfpages}
\usepackage{amsfonts,amssymb,amsmath}
\usepackage{graphicx}
\usepackage{epsfig,array,cite}
\usepackage{epsf,exscale,times}
\usepackage{color,multicol}

\let \bd = \textbf
\let \mbd = \mathbf
\let \it = \textit
\let \t  = \text

\def \p1#1{#1^{-1}}

\def \~#1{\tilde{#1}}

\let \nn = \nonumber

\begin{document}
\title{Dual-Function MIMO Radar Communications System Design Via Sparse Array Optimization}
\author{Xiangrong Wang, Aboulnasr Hassanien, and Moeness~G.~Amin
\thanks{Xiangrong Wang is with School of Electronic and Information Engineering, Beihang University, Beijing, China, 100191. E-mail: xrwang@buaa.edu.cn.
Aboulnasr Hassanien is with the Department of Electrical Engineering, Wright State University, Dayton, OH~45435, USA. E-mail: hassanien@ieee.org. Moeness Amin is with the Center for Advanced communications, Villanova University, PA~19085, USA. E-mail: moeness.amin@villanova.edu.
}
\thanks{The work by X Wang is supported by National Natural Science Foundation of China under Grant No. 61701016. The work by Dr M Amin is supported by the National Science Foundation under Grant No. 1547420.}
}
\maketitle

\vspace{-0.5cm}
\begin{abstract}

Spectrum congestion and competition over frequency bandwidth could be alleviated by deploying dual-function radar-communications systems, where the radar platform presents itself as a system of opportunity to secondary communication functions. In this paper, we propose a new technique for communication information embedding into the emission of multiple-input multiple-output (MIMO) radar using sparse antenna array configurations. The phases induced by antenna displacements in a sensor array are unique, which makes array configuration feasible for symbol embedding. We also exploit the fact that in a MIMO radar system, the association of independent waveforms with the transmit antennas can change over different pulse repetition periods without impacting the radar functionality. We show that by reconfiguring sparse transmit array through antenna selection and reordering waveform-antenna paring, a data rate of megabits per second can be achieved for a moderate number of transmit antennas. To counteract practical implementation issues, we propose a regularized antenna selection based signaling scheme. The possible data rate is analyzed and the symbol/bit error rates are derived. Simulation examples are provided for performance evaluations and to demonstrate the effectiveness of proposed DFRC techniques.

\end{abstract}


\IEEEpeerreviewmaketitle

\section{Introduction}

Recently, the ongoing intensive research has been developing multi-function solutions to the coexistence of radar and communications in the increasingly congested radio frequency (RF) spectral environment. Competition over frequency spectrum between radar and communications could be significantly alleviated when both systems are allowed to share the same spectrum resources and a single platform hardware \cite{Griffiths2013, Bliss2014, Hayvaci2014, Baylis2014}. This requires the establishment of dual system functionalities where identical signals, same frequency and bandwidth, and a common transmit platform are deployed to fulfill the objectives of both radar and communication operations \cite{Huang2015, Griffiths2015}. A dual-function radar-communications (DFRC) system utilizing waveform diversity in tandem with amplitude/phase control has been introduced in a number of papers \cite{Blunt2010, Ciuonzo2015, Guerci2015, Aubry2014, Aubry2015,Li2016}. It is assumed that the primary function of the antenna array is to enable a pulsed radar emission while providing the signal and system of opportunity to a secondary communication function concurrently during the radar pulse and with the same bandwidth. The DFRC systems are capable of making full use of the radar resources such as high quality hardware and high transmit power.

Different signaling schemes for embedding information into the radar pulsed emissions have been developed to establish a dual-function system that simultaneously performs both radar and communication functions \cite{Khawar2015, Wang2008, Song2010, Monte2015, Jakabosky2015, sahin2017, McCormick2017}. For example, sidelobe amplitude modulation, coherent and noncoherent phase modulation, multi-waveform amplitude-shift keying (ASK) were proposed to successfully embed information into the radar emission \cite{Hassanien2015, Hassanien2015a,Hassanien2016,Hassanien2016a}. However, all these signaling strategies were proposed for information embedding into the traditional phased array radar, where only scaled versions of a single waveform are transmitted, and thus cannot exploit waveform diversity. The multiple-input multiple-output (MIMO) radar generates a set of orthogonal waveforms via each element, thereby resulting in waveform diversity. For DFRC systems, the waveforms, that are simultaneously emitted from an antenna array in a MIMO arrangement, combine in the far-field to realize a desired radar waveform in one spatial direction and an information-bearing communication signal in another direction \cite{Skolnik1962, Li2007, Li2008}. Information embedding into the emission of MIMO radars has been considered in \cite{Hassanien2016b}, where one phase-shift keying (PSK) communication symbol is embedded in each orthogonal waveform. The achievable symbol rate is restricted by the number of orthogonal waveforms. To increase the data rate, frequency hopping codes were utilized to generate a set of orthogonal waveforms in \cite{Hassanien2017}. All these work has assumed either a uniform linear array (ULA) or an arbitrarily shaped array. To the authors' best knowledge, not much effort has been exerted to fully deploy spatial degrees for the design of dual-functional systems.

The successful co-existence of radar and communication functions requires not only the temporal diversity provided by advances in orthogonal waveform design, but also the spatial degrees of freedom (DoFs) brought about by properly utilizing the multi-sensor transmit/receive array configurations for suppressing cross-interference between the two functions. Although the nominal array configuration for existing DFRC systems is uniform and fixed-structured, it is not necessarily optimum in every sense and completely ignores the additional DoFs offered by array configurations. As antenna array technology progresses, sophisticated antenna selection schemes through RF switches and array reconfiguration methods that were previously infeasible begin to become possible. Sparse antenna arrays with non-uniform inter-element spacing attract increased attention in multi-sensor transmit/receive systems as an effective solution to reduce the system's complexity and cost, yet retain multifaceted benefits \cite{Amin2016,Wang2014}. Taking the notion of sparse arrays further, here we propose a technique utilizing array configuration for reliable communication symbol embedding concurrently with MIMO radar operation through antenna selection. We investigate the problem of expressing sparse array configurations and their association with independent waveforms as unique communication symbols. In spectrum sharing perspective, the deployment of reconfigurable sparse arrays by antenna selection can, undoubtedly, alleviate pressures on the resource management and efficiency requirements. Simulation results show that the versatility of sparse array configurations facilitates the realization of multiple functions on the same system.

The novelty of this paper is summarized as follows:
\begin{itemize}
\item
We propose an antenna selection based signaling strategy for DFRC systems to embed communication symbols into transmit array configurations.
\item 
We propose a hybrid selection and permutation strategy to combine array reconfiguration with waveform-antenna paring for communication symbol embedding in MIMO radars, which can achieve a high data rate and significantly reduce symbol error rate.
\item 
From the viewpoint of practical hardware implementation, we propose a regularized antenna selection based modulation scheme for DFRC systems, which is capable of achieving the bit error rate (BER) as low as binary PSK (BPSK) and high robustness against communication angle estimation error.

\end{itemize}

The rest of the paper is organized as follows. We provide the system configuration and signal model of the DFRC system with antenna selection network in section \ref{sec:model}. The unrestricted antenna selection based signaling strategy is proposed in section \ref{sec:anselect}. We then combine array reconfiguration with reordering waveform-antenna paring for high data rate communications in section \ref{sec:hybrid}. The regularized antenna selection scheme is elaborated in section \ref{sec:regularized}. Simulation results are provided in section \ref{sec:simulations}. Section \ref{sec:conclusions} summarizes the work of this paper.

\section{System Configuration and Signal Model}
\label{sec:model}

\begin{figure}[!ht]
  \centering
    \includegraphics[trim = {6cm 2cm 6cm 3cm}, scale=0.45]{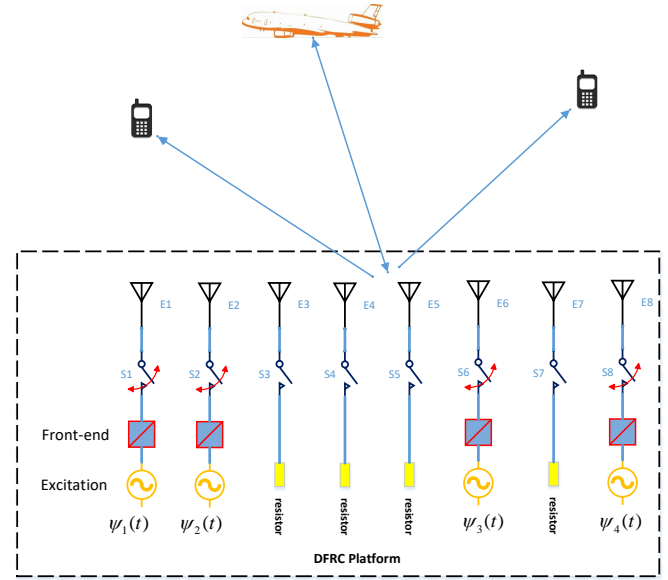}
    \caption{Joint platform of a DFRC system with antenna selection network.}
\label{fig_1}
\end{figure}

We consider a joint MIMO-radar communications platform equipped with a reconfigurable transmit antenna array through an antenna selection network as shown in Fig. \ref{fig_1}. This joint system can simultaneously detect radar targets of interest while sending communication symbols to downlink users. There are $M$ transmit antennas uniformly located in the platform with an inter-element spacing of $d$ and $K$ ($K\!\!<\!\!M$) front-ends installed for waveform transmitting. The antenna selection network comprises $M$ RF switches and their on/off status can be changed to connect/disconnect the corresponding antennas with the following front-ends. Note that only $K$ antennas are switched on for waveform transmitting during each pulse repetition interval (PRI) and the remaining $M-K$ antennas are either switched off or connected to resistors. Suppose a transmit array is configured with $K$ selected antennas located at $p_kd, k=1,\ldots,K$ with $p_k \in \{0,\ldots, M-1\}$. The radar receiver employs an array of $N$ receive antennas with an arbitrary linear configuration. It is assumed that both the transmit and receive arrays are closely spaced such that a target in the far-field would be seen from the same direction by both arrays.  Without loss of generality, a single-element communication receiver is assumed to be located in direction $\theta_c$, which is exactly known to the transmitter.

Let $\Psi_k(t), k=1, \ldots, K$ be a set of $K$ orthogonal waveforms, each occupying the same bandwidth. In other words, the spectral contents of all waveforms overlap in the frequency domain. Assume that each waveform is normalized to have unit power, i.e., $\int_{T} |\Psi_k(t)|^2 dt = 1$, with $T$ and $t$ denoting the waveform duration and the fast time index, respectively. It is further assumed that the orthogonality condition $\int_T \Psi_k(t) \Psi_{k'}^*(t) dt = 0$ is satisfied for $k \neq k'$, where $()^*$ stands for the complex conjugate. Assume that $Q$ far-field targets of interest arriving from the directions $\theta_q, q = 1, \ldots, Q$, located within the radar main beam, are observed in the background of strong clutter and interferences, such as television, radio and signals from other commercial communication services as well as deliberate jammers. The $N\times 1$ baseband representation of the signals at the output of the radar receive antenna array is given by, 
\begin{equation}
\label{eq:expre_xt}
\bd x(t;\tau) = \sum_{q=1}^Q \beta_q(\tau) \left(\tilde{\bd a}^T(\theta_q) \mbd{\Psi}(t) \right) \bd b(\theta_q) + \bd n(t;\tau),
\end{equation}
where $\tau$ is the pulse number, $\beta_q(\tau)$ is the $q$th target reflection coefficient\footnote{The target reflection coefficients are assumed to obey the Swerling-II target model \cite{Hassanien2016}, i.e., they remain constant during the entire pulse duration, but vary independently from pulse to pulse},  $\tilde{\bd a}^T(\theta_q)$ and $\bd b(\theta_q)$ denote the steering vectors of the sparse transmit array and the receive array, respectively, $(\cdot)^T$ stands for the transpose,  $\mbd{\Psi}(t) \triangleq [\Psi_1(t), \ldots, \Psi_K(t)]^T$ is the $K \times 1$ vector of orthogonal waveforms, and $\bd n(t; \tau)$ is the $N \times 1$ vector of zero mean summarizing the unwanted clutter, interferences and white noise during the $\tau$th radar pulse. In \eqref{eq:expre_xt}, the sparse transmit array steering vector, $\tilde{\bd a}(\theta)$, can be defined as,
\begin{equation}
\tilde{\bd a}(\theta) = [e^{jk_0 p_1 d \sin\theta}, \ldots, e^{jk_0 p_K d \sin\theta}]^T,
\end{equation}
where $k_0=2\pi /\lambda$ is the wavenumber and $p_k \in \{0, \ldots, M-1\}, k=1, \ldots, K$. The  steering vector of the MIMO radar receive array, $\bd b(\theta)$, can be defined in a similar way as that of $\tilde{\bd a}(\theta)$.

The signal at the output of the communication receiver can be modelled as
\begin{equation}
\label{eq:expre_commu}
x_c(t;\tau) = \alpha_{\rm ch}(\tau) \tilde{\bd a}^T(\theta_c) \mbd{\Psi}(t) + n_c(t;\tau),
\end{equation}
where $\alpha_{\rm ch}(\tau)$ is the channel coefficient of the received signal that summarizes the propagation environment between the transmit array and the communication receiver during the $\tau$th pulse and $\tilde{\bd a}(\theta_c)$ is the steering vector of the selected transmit array toward the communication direction $\theta_c$. In addition, $n_c(t;\tau)$ is the noise signal interfering the communication process in the $\tau$th radar pulse. We assume that the channel coefficient $\alpha_{\rm ch}$ is known or accurately estimated and remains unchanged during the entire coherent processing interval of the DFRC system. Therefore, for the rest of this paper, we remove the dependency of the channel coefficient on the pulse index $\tau$.

At the MIMO radar receiver, the received signal components associated with the individual transmitted waveforms can be obtained using matched filtering to Eq. (\ref{eq:expre_xt}). The signals observed at the output of the radar receiver are the $KN\times 1$ extended vector of virtual data, that is,
\begin{eqnarray}
\label{eq:expre_radar}
\bd y_r(\tau) &=& \t{vec}\left( \int_T \bd x(t;\tau) \mbd{\Psi}^H(t) dt \right) , \\
                &=& \sum_{q=1}^Q \beta_q(\tau) [\tilde{\bd a}(\theta_q) \otimes \bd b(\theta_q)] + \bd n(\tau), \nn
\end{eqnarray}
where $\t{vec}(\cdot)$ is the operator that stacks the columns of a matrix into one column vector, $\otimes$ denotes the Kronecker product, $^H$ stands for the Hermitian transpose, and $\bd n(\tau) = \t{vec}\left( \int_T \bd n(t;\tau) \mbd{\Psi}^H(t) dt \right)$ is the $KN \times 1$ additive noise term after matched filtering.

The communication receiver is assumed to have perfect knowledge of the orthogonal waveforms $\Psi_k(t), k=1, \ldots, K$. Moreover, it is assumed that the phase synchronization between the transmit array and the communication receiver is adjusted. Matched filtering the received data in Eq. (\ref{eq:expre_commu}) to each waveform $\Psi_k(t)$ yields,
\begin{align}
\label{eq:expre_ycm}
y_{c,k}(\tau) &= \int_T  x_c(t;\tau) \Psi_k^*(t) dt, \\
              &= \alpha_{\rm ch} \tilde{\bd a}_k(\theta_c) + n_{c,k}(\tau), \quad k=1, \ldots, K \nn
\end{align}
where $\tilde{\bd a}_k=e^{jk_0p_k d \sin\theta_c}$ denotes the $k$th entry of the selected transmit array steering vector $\tilde{\bd a}(\theta_c)$ and $n_{c,k}(\tau) = \int_T n_c(t;\tau) \Psi_k^*(t) dt, k=1, \ldots, K$ are additive noise terms after matched filtering. Array configurations denote the spatial DoFs and can be combined with waveform design in temporal domain to embed communication symbols concurrently with MIMO radar functions. We elaborate on the deployment of sparse arrays for symbol embedding in DFRC systems in Sections~\ref{sec:anselect}-\ref{sec:regularized}.

\section{Antenna Selection Based Signaling Strategy for DFRC Systems}
\label{sec:anselect}

There are totally $M$ antennas installed in the common transmit platform and an antenna selection network is deployed to select $K$ out of $M$ antennas. We deploy $K$ orthogonal waveforms, $\Psi_1(t), \ldots, \Psi_K(t)$, and transmit them via the selected $K$ antennas. It is clear that the steering vectors of selected sparse arrays can be estimated by the communication receiver after matched filtering and utilized to embed communication symbols from Eq. (\ref{eq:expre_ycm}). 

\subsection{Information Embedding Scheme}

The steering vector of the $M$-antenna full transmit array is denoted as $\bd a(\theta)$ and can be expressed as
\begin{equation}
\label{eq:expre_a}
\bd a(\theta) = [e^{jk_0 p_1 d \sin\theta}, \ldots, e^{jk_0 p_M d \sin\theta}]^T.
\end{equation}
Denote the $K \times M$ selection matrix during the $\tau$th radar pulse as $\bd P(\tau) \in \{0,1\}^{K \times M}$, where there is only one entry being ``1'' in each row and in the $k$th column corresponding to the $k$th selected antenna, $k \in \{1, \ldots, M\}$. Applying the selection matrix $\bd P(\tau)$ to the steering vector $\bd a(\theta)$ of the full transmit array yields the $K \times 1$ steering vector of the selected subarray, that is,
\begin{equation}
\label{eq:expre_aselected}
\tilde{\bd a}(\theta;\tau) = \bd P(\tau) \bd a(\theta).
\end{equation}
The $K$ orthogonal waveforms are transmitted via the selected $K$ antennas, and the $N \times 1$ complex vector of the radar received observations in Eq. (\ref{eq:expre_xt}) can be rewritten as,
\begin{equation}
\label{eq:expre_xtt}
\bd x(t,\tau) = \sum_{q=1}^Q \beta_q(\tau) [\bd a^T(\theta_q) \bd P^T(\tau) \mbd{\Psi}(t)]  \bd b(\theta_q) + \bd n(t,\tau).
\end{equation}

Let $\mathcal{P}=\{1, \ldots, M\}$ label the full set of antennas installed in the transmit platform. During each radar pulse, a subset $\mathcal{S}$ of $K$ antennas are selected from the full set $\mathcal{P}$ for waveform transmitting. Such a selection is essentially a combinatorial problem. There are totally $L=C_M^K=\frac{M!}{K!(M-K)!}$ different subsets, $\mathcal{S}_l \subset \mathcal{P}, l=1, \ldots, L$, and each subset $\mathcal{S}_l$ corresponds to a unique selection matrix $\bd P(\tau)$ with ``1'' entries located in the columns indicated by $\mathcal{S}_l$, which in turn corresponds to a unique steering vector $\tilde{\bd a}(\theta;\tau)$. For each subarray $\mathcal{S}_l$, a communication symbol consisting of $N_b$ bits can be defined. 

Assume that the communication receiver knows its direction $\theta_c$ relative to the stationary MIMO transmit platform. The signal at the output of the communication receiver antenna in Eq. (\ref{eq:expre_commu}) is remodelled as,
\begin{eqnarray}
x_c(t,\tau) = \alpha_{\rm ch} \bd a^T(\theta_c) \bd P^T(\tau) \mbd{\Psi}(t) + n_c(t,\tau), 
\end{eqnarray}
Matched filtering the received data with the set of $K$ orthogonal waveforms yields,
\begin{eqnarray}
\label{eq:expre_y_cselection}
\bd y_c(\tau) &=& \t{vec} \left\lbrace \int_T x_c(t,\tau) \mbd{\Psi}^H(t) dt \right\rbrace, \\
& = & \alpha_{\rm ch} \bd P(\tau) \bd a(\theta_c) + \bd n_c(\tau), \nn\\
&=& \alpha_{\rm ch} \tilde{\bd a}(\theta_c;\tau) + \bd n_c(\tau). \nn
\end{eqnarray}
where $\bd y_c(\tau)=[y_{c,1}(\tau), \ldots, y_{c,K}(\tau)]^T$ and $\bd n_c(\tau)=[n_{c,1}(\tau), \ldots, n_{c,K}(\tau)]^T$. Thus, the communication receiver signal at the output of the matched-filter is a scaled and noisy selection of the full steering vector $\bd a(\theta_c)$, meaning that the selected sparse array $\mathcal{S}_l$ can be recovered from the received vector $\bd y_c(\tau)$. We propose to utilize the steering vector of the sparse transmit array $\mathcal{S}_l$ as codes to embed communication symbols. 

\subsection{Detection of communication Symbols}

As mentioned above, the steering vector of the selected sparse array $\mathcal{S}_l, l=1, \ldots, L$ can be utilized to embed communication symbols. Thus, a dictionary of $L$ unique symbols can be constructed as
\begin{equation}
\label{eq:expre_dictionary}
\mathbb{D} = \{\tilde{\bd a}_1(\theta_c), \ldots, \tilde{\bd a}_L(\theta_c)\},
\end{equation}
where $\tilde{\bd a}_l(\theta_c)$ is the steering vector corresponding to the selected sparse array $\mathcal{S}_l\subset \mathcal{P}$ towards the communication receiver. The cardinality $L$ of the dictionary determines the capacity of the communication system and is usually not a power of 2. The fact that only a subset of $L$ available symbols is required offers flexibility in the design of the actual system as well as improved noise immunity. A compressive study of the selection of symbol subset will be discussed in the section \ref{subsec:selection}.

Let us assume that the channel is estimated accurately. In practice, training sequences can be periodically transmitted to update the channel estimate and adjust phase synchronization between the transmit array and the communication receiver. During each radar pulse, an $N_b$-bit communication information is first converted into the corresponding decimal number $n_d$. The $K$ antennas comprised by the sparse array $\mathcal{S}_{n_d}$ are then selected to transmit the $K$ orthogonal waveforms. The steering vector of the selected transmit sparse array during the $\tau$th radar pulse can be estimated as,
\begin{equation}
\hat{\bd a}(\theta_c;\tau) = (1/\alpha_{\rm ch}) \bd y_{c}(\tau).
\end{equation}
The communication receiver then calculates the distance between the estimated vector $\hat{\bd a}(\theta_c;\tau)$ and each element of the dictionary $\mathbb{D}$, that is $D^l = \|\hat{\bd a}(\theta_c;\tau) - \tilde{\bd a}_l(\theta_c)\|_2, l=1, \ldots, L$. The embedded communication symbol can be found with the smallest distance $\min_l D^l$ and then converted into the corresponding binary sequence. The detection of each communication symbol requires $LK$ complex multiplications, thus computational complexity increases proportionally with the data rate.

Given that the radar transmits one symbol per pulse, the symbol rate of the communication system is identical to the pulse repetition frequency ($f_{\t{PRF}}$).  The number of bits that can be transmitted per symbol is
\begin{equation}
N_b = \lfloor \t{log}_2 L \rfloor,
\end{equation}
where $\lfloor \cdot \rfloor$ stands for the largest integer that is no greater than the argument. Thus, the resulting data rate of the antenna selection based dual-functional systems is $N_b \times f_{\t{PRF}}$ bit per second (bps).

\subsection{Angular Ambiguities}
\label{subsec:ambiguity}

Note that the steering vector comprises the phase terms $\phi_k = k_0p_kd\sin\theta_c, k=1, \ldots, K, p_k \in {0, \ldots, M-1}$ produced by the displacement of the $K$ selected antennas. The mapping between the phase term $\phi_m=k_0p_m d \sin\theta_c, m=1, \ldots, M$ and the antenna position $p_m$ is not one-to-one, as the phase is periodic with a period of $2\pi$. That means there may exist multiple antennas in the array producing the same phase term, giving rise to the problem of angular ambiguity. The trivial ambiguity happens when $\theta_c=0$, that is when the communication receiver is at broadside direction. The phases are zero regardless of antenna positions and all entries of $\bd a(\theta_c)$ are one. Thus, no information can be embedded via antenna selection. When $|\theta_c|$ is small, the phases produced by all the $M$ antennas are equally spaced on the unit circle with values $k_0md\sin\theta_c, m=0, \ldots, M-1$. As $\theta_c$ increases, the phase difference between two adjacent antennas increases and the $M$th antenna reaches an angle for which we have the largest spread around the unit circle, that is,
\begin{equation}
\label{eq:expre_thetacmax}
\phi_M=k_0(M-1)d\sin\theta_c=2\pi-k_0d\sin\theta_c.
\end{equation}
Solving Eq. (\ref{eq:expre_thetacmax}) yields $\theta_{cm}=\t{sin}^{-1}\left( \frac{2\pi}{Mk_0d} \right)$. We refer to this angle as the maximal spread angle. When $\theta_c > \theta_{cm}$, it is likely that two or more antennas exhibit the same phase value. This happens when their phases are equal modulo $2\pi$. In general, angular ambiguities happen when $(M-m)k_0d\sin\theta_c=2\pi$, giving $\theta_c=\t{sin}^{-1} \frac{2\pi}{(M-m)k_0d}$ for $m=1, \ldots, M-1$. It is important to note, however, that when the arrival angle of communication receiver $\theta_c$ is small, the performance may be poor if the full number of bits is used. Therefore, it is advantageous to transmit at the maximal spread angle $\theta_{cm}$ which gives the best performance. We then describe a scheme to mitigate the ambiguities and steer the performance of the maximal spread angle to any receiver spatial angle.

The ambiguities described above can be mitigated by introducing additional phase rotation to each transmit antenna. Denote the vector of phase rotations assigning to the $M$ antennas as $\bd u = [e^{j \varphi_1}, \ldots, e^{j \varphi_M}]^T$. We pre-multiply element-wise at the transmitter the vector of orthogonal waveforms, $\mbd{\Psi}(t)$, by the selected phase vector $\bd P(\tau)\bd u$. That means once an antenna is selected, the corresponding phase rotation is multiplied. Then the vector of phase-shifted waveforms become $\tilde{\mbd{\Psi}}(t) = \t{diag}(\bd P(\tau) \bd u) \mbd{\Psi}(t)$, with $\t{diag}(\cdot)$ denoting a diagonal matrix with the vector $\cdot$ populating along the diagonal. The set of rotated waveforms $\tilde{\mbd{\Psi}}(t)$ still preserve the orthogonality, which is proved as follows,
\begin{eqnarray}
&& \int_0^T \tilde{\mbd{\Psi}}(t) \tilde{\mbd{\Psi}}^H(t) dt \\
&=&  \t{diag}(\bd P(\tau) \bd u) \int_0^T \mbd{\Psi}(t) \mbd{\Psi}^H(t)dt \; \t{diag}(\bd P(\tau) \bd u)^H , \nn\\
&=& \t{diag}(\bd P(\tau) \bd u) \t{diag}(\bd P(\tau) \bd u^*), \nn\\
&=&  \bd I, \nn
\end{eqnarray}
Thus, the phase-rotated waveforms $\tilde{\mbd{\Psi}}(t)$ does not affect the normal operation of radar functions. The matched-filtered signal at the communication receiver in Eq. (\ref{eq:expre_y_cselection}) becomes, 
\begin{equation}
\bd y_c(\tau) = \alpha_{\rm ch} \t{diag}(\bd P(\tau)\bd u) \tilde{\bd a}(\theta_c;\tau) + \bd n_c(\tau).
\end{equation}
The received signal vector now has phases $\tilde{\phi}_k = \phi_k + \varphi_k = k_0 p_k d \sin\theta_c + \varphi_{i_k}, k=1,\ldots,K, i_k \in \{1, \ldots, M\}$. Thus, we can deduce a specific phase rotation for each transmit antenna, such that the phases of all $M$ antennas are uniformly distributed around the unit circle at the spatial angle $\theta_c$ of the communication receiver. That means, $\tilde{\phi}_m=2\pi (m-1)/M, m=1, \ldots, M$. Then, the phase rotation for the $m$th antenna can be calculated as,
\begin{equation}
\label{eq:expre_varphiik}
\varphi_m = \frac{2\pi(m-1)}{M} - k_0p_md \sin\theta_c, m=1, \ldots, M.
\end{equation}
In this manner, not only are we able to mitigate the ambiguities, but also to deliver the best symbol dictionary to any receiver.

\subsection{Symbol Error Rate}

Let us assume without loss of generality that the transmitted sparse array is $\mathcal{S}_i$, whose corresponding steering vector $\tilde{\bd a}_i$ comprises $K$ phases of value $\tilde{\phi}_k=2\pi(i_k-1)/M, i_k=\{1, \ldots, M\}$. It is worth noting that the dependence of the steering vector $\tilde{\bd a}_i$ on the angle $\theta_c$ of the communication receiver is suppressed due to the additional phase rotations. The symbols in the dictionary defined in Eq. (\ref{eq:expre_dictionary}) 
%
%
change to
\begin{equation}
\tilde{\bd a}_l = [e^{j 2\pi (l_1-1)/M}, \ldots, e^{j 2\pi (l_K-1)/M}]^T, l_k =\{1, \ldots, M\}.
\end{equation}
Let us define the distance between the estimated steering vector $\hat{\bd a}_i$ and each code $\tilde{\bd a}_l$ in the dictionary as $D^l = \|\hat{\bd a}_i-\tilde{\bd a}_l\|_2$. Then the probability of a correct symbol detection is given by
\begin{equation}
P_d = P\left( D^i < D^l, \forall l = 1, \ldots, L, l \neq i \right). 
\end{equation}
It is worth noting that a symbol error may not occur even if the noise places some phase $\tilde{\phi}_k=2\pi(i_k-1)/M$ of the received signal closer to another constellation, $2\pi(l_k-1)/M, i_k, l_k \in \{1, \ldots, M\}$, such that $l_k \neq i_k, k=1, \ldots, K$, provided that $D^i < D^l$. Thus, for each symbol $\mathcal{S}_i$, we have
\begin{equation}
\label{eq:expre_PDiDl}
P\left( D^i < D^l \right) \geq \Pi_{k=1}^K P(D^i_k < D^l_k),
\end{equation}
where $P(D^i_k < D^l_k)$ denotes the probability of a correct detection of the $k$th phase term. Detecting each phase of the steering vector is similar to the M-ray phase-shift keying (M-ary PSK) scenario, where every phase $\tilde{\phi}_k$ is taken out of $M$ uniformly distributed signal constellations around a unit circle with an angular separation of $\gamma = 2\pi/M$. The average probability of symbol error for M-ary PSK modulation with sufficiently high signal-to-noise ratio (SNR) is \cite{Haykin2008},
\begin{equation}
Q(\rho, \gamma) = \t{erfc} \left( \sqrt{\rho} \sin(\frac{\gamma}{2}) \right),
\end{equation}
where $\rho$ stands for the SNR and $\t{erfc}$ denotes the complementary error function. Thus, we have that
\begin{equation}
\label{eq:expre_Pdi}
P\left( D_k^i < D_k^l \right) = 1- Q(\rho, \gamma), k=1, \ldots, K.
\end{equation}
Substituting Eq. (\ref{eq:expre_Pdi}) into Eq. (\ref{eq:expre_PDiDl}) yields the lower bound of the detection probability,
\begin{equation}
\label{eq:expre_Pd1}
P_d \geq [1- Q(\rho, \gamma)]^K.
\end{equation}
The upper bound of symbol error rate (SER) is then obtained by,
\begin{equation}
\label{eq:expre_pe1}
P_e = 1-P_d \leq 1 - [1- Q(\rho, \gamma)]^K.
\end{equation}
The embedded symbol is detected by comparing the distance between the estimated steering vector and each code in the dictionary. Thus, it is preferred that the distance between any two code vectors in the dictionary be maximized. For radar modalities, target detection is the main objective. The detection performance is directly related to the efficacy of clutter cancellation. As the same set of $K$ orthogonal waveforms $\mbd{\Psi}(t)$ are transmitted during each PRI, there is no attendant Doppler coherency degradation as existed in waveform modulation scheme proposed in \cite{Blunt2010a}. However, the transmit array configuration affects the radar detection performance significantly. Thereby, the selection of symbol subset should consider two criteria together, the performance of communication functions and a satisfying radar transmit beampattern.

\subsection{Selection of Constellation Symbols}
\label{subsec:selection}

We consider the first criterion of communication performance, that is selecting a subset of $L_b=2^{N_b}$ symbols from $L$ candidates, such that the distance between any two symbols in the dictionary is maximized. Without loss of generality, the total number $M$ of installed antennas is assumed to be even. As all the $M$ antennas are uniformly distributed around the unit circle with a phase difference of $2\pi/M$, it is intuitive that the two symbols with the largest distance are $\tilde{\bd a}_1 = \{1, e^{j2\pi/M}, \ldots, e^{j2\pi(K-1)/M}\}$ and $\tilde{\bd a}_2 = \{-1, e^{j2\pi(M/2+1)/M}, \ldots, e^{j2\pi(M/2+K-1)/M}\}$. That means each pair of antennas in symbols $\tilde{\bd a}_1$ and $\tilde{\bd a}_2$ are center-symmetrically distributed in the upper and lower half circles, respectively. The largest distance can be calculated as $\|\tilde{\bd a}_1-\tilde{\bd a}_2\|^2 = 4K$. Initialize the symbol subset as $\mathbb{D}_c=\{\tilde{\bd a}_1, \tilde{\bd a}_2\}$, and $\bd z_l$ are selection vectors corresponding to $\tilde{\bd a}_l$ such that $\tilde{\bd a}_l=\bd a(\mathbb{Z}(\bd z_l)), l=1,2$, where $\mathbb{Z}(\bd z)$ denotes the sparse support of vector $\bd z$. The remaining $L_b-2$ symbols can be found as follows: 
\begin{eqnarray}
\label{eq:expre_subsetcomm}
\max_{\bd z, \nu} && \nu, \\
\t{subject to} && \|\t{diag}(\bd z) \bd a - \t{diag}(\bd z_l) \bd a\|_2^2 \geq \nu, l=1, \ldots, |\mathbb{D}_c|, \nn\\
               && \bd z \in \{0, 1\}^M, \; \bd 1^T \bd z = K, \nn
\end{eqnarray}
where $\bd a=[1, e^{j2\pi/M}, \ldots, e^{j2\pi(M-1)/M}]^T$ is the steering vector of the full array after phase rotation and the vector comprised of the non-zero entries of $\t{diag}(\bd z_l) \bd a$ is a symbol in $\mathbb{D}_c$. The selection variable $\bd z$ is binary with entry one denoting the corresponding antenna selected and entry zero discarded, $|\mathbb{D}_c|$ stands for the cardinality of the symbol subset $\mathbb{D}_c$. The constraint $\bd 1^T \bd z = K$ controls the number of selected antennas to be exactly $K$. 

As explained in \cite{Wang2014a}, the binary property of the selection variable $\bd z \in \{0, 1\}^M$ is tantamount to 
\begin{equation}
\label{eq:expre_binary}
\max_{\bd z} \; \bd z^T(\bd z-\bd 1) \; \t{subject to} \; 0 \leq \bd z \leq 1,
\end{equation}
where the inequality is the constraint applying to each entry in vector $\bd z$. Combining Eqs.~(\ref{eq:expre_binary}) with (\ref{eq:expre_subsetcomm}) yields the following formulation,
\begin{eqnarray}
\label{eq:expre_subsetcomm1}
\max_{\bd z, \nu} && \nu + \mu[\bd z^T(\bd z-\bd 1)], \\
\t{subject to} && \bd a^H \t{diag}(\bd z) \bd a-2\bd a^H \t{diag}(\bd z) \t{diag}(\bd z_l) \bd a+ K \leq \nu, \nn\\
&& l=1, \ldots, |\mathbb{D}_c|, \nn\\
&& 0 \leq \bd z \leq 1, \; \bd 1^T \bd z = K, \nn
\end{eqnarray}
where the trade-off parameter $\mu$ is used to control the emphasis between symbol distance and the boolean property of the selection vector $\bd z$. Equal importance can achieved by setting $\mu=1$.

Next, we consider the second criterion of radar transmit pattern synthesis and the optimum dictionary is denoted as $\mathbb{D}_r$. As shown in Eq. (\ref{eq:expre_radar}), the virtual extended signal vector of the MIMO radar is the Kronecker product between transmit and receive array steering vectors. That is,
\begin{equation}
\bd c(\theta) = \tilde{\bd a}(\theta) \otimes \bd b(\theta).
\end{equation}
Assume that the beamforming weight vector is $\bd w$, the overall beampattern of MIMO radar can be expressed as
\begin{equation}
B(\theta) = \left|\bd w^H \bd c(\theta)\right| = \left|\bd w^H [\tilde{\bd a}(\theta) \otimes \bd b(\theta)]\right|.
\end{equation}
We can see that the shape of overall beampattern is affected by both transmit and receive array configurations. Since the structure of receive array is fixed, it is preferred that sparse transmit array configurations satisfy a certain desired power radiation pattern, when combined with the given receive array. The main function of MIMO radar is to concentrate the transmit power within a certain angular sector $\Theta=[\theta_{\t{min}}, \theta_{\t{max}}]$, where the radar signal may come from. The beampattern corresponding to the sidelobe region $\bar{\Theta}$ is required to be less than a pre-defined sidelobe level $\epsilon$. The selection of symbol subset satisfying the criterion of radar function can be formulated as follows:
\begin{eqnarray}
\label{eq:expre_method2}
\max_{\bd z, \bd w, \rho} && \rho + \mu[\bd z^T(\bd z - \bd 1)], \\
\t{s.t.} && |\bd w^H \bd c(\theta_i) - e^{j\mu(\theta_i)}| \leq \rho, \; \theta_i \in \Theta, i = 1, \ldots, L_m, \nn\\
         && |\bd w^H \bd c(\theta_k)| \leq \epsilon, \; \theta_k \in \bar{\Theta}, k = 1, \ldots, L_s \nn\\
         && |\bd J_m\bd w| \leq z_m, m = 1, \ldots, M \nn\\
         && 0 \leq \bd z \leq 1, \bd 1^T \bd z = K, \nn
\end{eqnarray}
where $\theta_i, i=1,\ldots,L_m$ and $\theta_k, k=1,\ldots,L_s$ are $L_m$ and $L_s$ samples of the mainlobe region $\Theta$ and sidelobe region $\bar{\Theta}$, respectively, and $\mu(\theta)$ is the user-defined mainlobe phase profile, $\rho$ denotes the allowable maximum mainlobe ripple. The weight vector $\bd w$ exhibits a block sparsity with $N-K$ blocks of $M$ entries being zero. In addition, the matrix $\bd J_m \in \{0,1\}^{N \times MN}$ is utilized to extract the $[(m-1)N+1] \sim (mN)$th entries of the weight vector $\bd w$. The matrix has ``one'' entry in each row and in the $[(m-1)N+1] \sim (mN)$th columns, and all other entries being zero. The constraints $|\bd J_m\bd w| \leq z_m, m = 1, \ldots, M$ are used to promote the same group sparsity of weight vector $\bd w$ as the selection variable $\bd z$.

Clearly, the objective functions in Eqs. (\ref{eq:expre_subsetcomm1}) and (\ref{eq:expre_method2}) are concave, and it is difficult to maximize them directly. A sequential convex programming (SCP) based on iteratively linearizing the concave objective function is then utilized to reformualte the non-convex problem to a series of convex subproblems, each of which can be optimally solved using convex programming \cite{Boyd2004,Fazel2003}. Taking the problem in Eq. (\ref{eq:expre_method2}) as an example, the symbol selection in the (k+1)th iteration can be formulated based on the solution $\bd z^{(k)}$ from the kth iteration as,
\begin{eqnarray}
\label{eq:expre_subsetcomm2}
\max_{\bd z, \bd w, \rho} && \rho + \mu[(2\bd z^{(k)}-\bd 1)^T \bd z - \bd z^{(k)T} \bd z^{(k)}], \\
\t{s.t.} && |\bd w^H \bd c(\theta_i) - e^{j\mu(\theta_i)}| \leq \rho, \; \theta_i \in \Theta, i = 1, \ldots, L_m, \nn\\
         && |\bd w^H \bd c(\theta_k)| \leq \epsilon, \; \theta_k \in \bar{\Theta}, k = 1, \ldots, L_s \nn\\
         && |\bd J_m\bd w| \leq z_m, m = 1, \ldots, M \nn\\
         && 0 \leq \bd z \leq 1, \bd 1^T \bd z = K. \nn
\end{eqnarray}
Note that the SCP is a local heuristic and its performance depends on the initial point $\bd z^{(0)}$. It is, therefore, feasible to construct the symbol dictionary by initializing the algorithm with different feasible points $\bd z^{(0)}$. The symbol subset selection considering both criteria can be achieved by combining the constraints in Eqs. (\ref{eq:expre_subsetcomm1}) and (\ref{eq:expre_method2}). The detailed description of the symbol subset selection is illustrated in Table. \ref{table_1}. After obtaining the selection vector $\bd z^*$, the corresponding symbol can be calculated as $\tilde{\bd a}= \bd a(\mathbb{Z}(\bd z^*)) = \{e^{j2\pi (m-1)/M}, z^*_m=1,m=1, \ldots, M\}$.

\begin{table*}[!h]
\caption{The detailed description of symbol subset selection}
\label{table_1}
\begin{center}
\begin{tabular}{l|l}
\hline
\hline
Initialization & Initialize symbol subset $\mathbb{D}=\{\}$, initialize $\mu=1$. \\
Step 1 (Outer Loop) & \textbf{WHILE:} $|\mathbb{D}| < L_b$,\\
Step 2 & Set $k=0$ and maximum iteration number $K_m$. Randomly initialize $\bd z^{(k)}$.\\ 
Step 3 (Inner Loop) & \textbf{FOR} $k < K_m$\\
       & Solve Eq. (\ref{eq:expre_subsetcomm1}) or (\ref{eq:expre_method2}) using Matlab embedded software CVX, set k=k+1;\\
       & \textbf{END OF INNER LOOP}\\
Step 4 & If the obtained selection vector $\bd z$ is boolean and not included in $\mathbb{D}$, calculate the corresponding symbol $\tilde{\bd a}$,\\
       & set $\mathbb{D}=[\mathbb{D}, \tilde{\bd a}]$ and go to Step 1.\\
Step 5 & If the obtained selection vector $\bd z$ is not boolean, go to Step 2; \\
Step 6 & \textbf{END OF OUTER LOOP} \\
\hline
\hline
\end{tabular}
\end{center}
\end{table*}

\section{Hybrid Selection and Permutation based Signaling Strategy for DFRC Systems}
\label{sec:hybrid}

The MIMO radar receiver requires the knowledge of transmit waveform and transmit antenna pairing, and does not require pinning a specific waveform to a specific antenna \cite{BouDaher2016}. The flexibility of varying the selected transmit antennas across $K$ orthogonal waveforms over different PRIs can be exploited to embed a large constellation of symbols. For each selected $K$-antenna sparse array, the number of symbols that can be embedded is a factorial of the number of transmit antennas, that is $K!$. Thus, combining antenna selection with permutating $K$ independent waveforms to each selected antenna over one PRI, a data rate of megabits per second can be achieved by a moderate number of transmit antennas. Taking this notion further, we propose a hybrid selection and permutation based signaling strategy for DFRC systems in this section. Since permutations used to assign the antennas to the waveform set are known to the radar, the reordering enables restoring the coherent structure of the MIMO radar data, i.e. the primary MIMO radar operation is unaffected by the secondary communication function.

The structure of hybrid selection and permutation based signaling strategy remains the same as that of selection only method and is depicted in Fig. \ref{fig_1}. There are $M$ antennas installed on the platform and a specific subset of $K$ antennas associated with the communication symbols are switched on for transmitting independent waveforms during each radar pulse. Denote the $K \times M$ selection matrix and $K \times K$ permutation matrix as $\bd P(\tau)$ and $\bd Q(\tau)$, respectively. The signal at the output of the communication receiver antenna is remodelled as,
\begin{equation}
x_c(t,\tau) = \alpha_{\rm ch} \bd a^T(\theta_c) \bd P^T(\tau) \bd Q^T(\tau) \mbd{\Psi}(t) + n_c(t,\tau).
\end{equation}
Matched filtering the received data with the set of orthogonal waveforms yields,
\begin{equation}
\bd y_c(\tau) = \t{vec} \left\lbrace \int_T x_c(t,\tau) \mbd{\Psi}^H(t) dt \right\rbrace = \alpha_{\rm ch} \bd M(\tau) \bd a(\theta_c) + \bd n_c(\tau),
\end{equation}
where $\bd M(\tau)=\bd Q(\tau) \bd P(\tau)$. Thus, the communication receiver signal at the output of the matched-filter is a (scaled and noisy) selected permutation of the steering vector $\bd a(\theta_c)$, meaning that the product of selection and permutation matrices $\bd M(\tau)$ can be recovered from the received vector $\bd y_c(\tau)$ by determining the ordering of $K$ selected transmit antennas. We propose to utilize the selected permutation of the steering vector $\bd M(\tau) \bd a(\theta_c)$, that is the ordered set of phases induced by selected antenna positions, as the codes to embed communication symbols.

To mitigate angular ambiguity and maximize communication performance, the phase rotation imposed to each transmit antenna per Eq. (\ref{eq:expre_varphiik}) can be deployed here. Thus, a dictionary of $K!\times L$ symbols is constructed as,
\begin{equation}
\mathbb{D} = \{\bd A_1, \ldots, \bd A_L\},
\end{equation}
where $\bd A_l=[\bd Q_1 \tilde{\bd a}_l, \ldots, \bd Q_{K!} \tilde{\bd a}_l]$ with $\bd Q_k, k=1, \ldots, K!$ denoting the permutation matrix. In addition, $\tilde{\bd a}_l=[e^{j\tilde{\phi}_{l_1}}, \ldots, e^{j\tilde{\phi}_{l_K}}]^T$ with $l_k \in \{1, \ldots, M\}$ and $\tilde{\phi}_{l_k} = 2\pi(l_k-1)/M$. During each radar pulse, the $K$ orthogonal waveforms $\Psi_k(t), k=1, \ldots, K$ are transmitted through the ordered subset of antennas with positions $p_k$ corresponding to the $N_b$-bit information. Assume that communication receiver has a prior knowledge of its angle $\theta_c$ relative to the joint transmit array. The ordered selected steering vector can be estimated as,
\begin{equation}
\hat{\bd a}(\theta_c;\tau) = (1/\alpha_{\rm ch})\bd y_c(\tau) \approx \bd M(\tau) \bd a(\theta_c).
\end{equation}
The communication receiver can then compare the estimated vector $\hat{\bd a}(\theta;\tau)$ to the dictionary $\mathbb{D}$ to obtain the embedded communication symbols. As there are $K!$ different ordering for each selected subarray, the message bits that can be transmitted during each pulse are
\begin{equation}
N_b = \lfloor \t{log}_2 (L \times K!) \rfloor = \lfloor \t{log}_2 L + \t{log}_2 K! \rfloor.
\end{equation}
Thus, the data rate, measured in bps, for the proposed hybrid selection and permutation based signaling scheme can be expressed as,
\begin{equation}
R = \lfloor \t{log}_2 L + \t{log}_2 K! \rfloor \times f_{\t{PRF}}.
\end{equation}

It is worth noting that, not only the data rate can be increased, but also the symbol error rate for the hybrid scheme can be significantly reduced compared with that of selection-only signaling scheme. The reason is that the permutation of antenna positions can be utilized to further increase the distance between the selected symbols $\tilde{\bd a}_l, l=1, \ldots, L_b$ in $\mathbb{D}_c$. The symbol selection for the dictionary $\mathbb{D}_p$ utilizing both antenna selection and permutation can be formulated as,
\begin{eqnarray}
\label{eq:expre_symbolpermu}
\max_{\bd Q, \bd P, \nu} && \nu, \\
\t{subject \; to}  &&  \|\bd Q \bd P \bd a - \bar{\bd a}_k \|^2 \geq \nu, k = 1, \ldots, |\mathbb{D}_p|, \nn
\end{eqnarray}
where $\bar{\bd a}_k = \bd Q_k \bd P_k \bd a \in \mathbb{D}_p$ are already-selected symbols. As the optimization variables $\bd Q$ and $\bd P$ are required to satisfy the conditions of permutation matrix and selection matrix, respectively, the problem is highly non-convex. Moreover, enumerating all $L \times K!$ different permutations and combinations is prohibitively exhausitive. Instead, we resort to enumerate the optimum permutation for each selected symbol in $\mathbb{D}_c$ and obtain a sub-optimum dictionary $\mathbb{D}_p$. It is worth noting that the proposed signaling modulation strategy is different from the waveform shuffling scheme introduced in \cite{BouDaher2016}, where permutation matrix $\bd Q$ only is utilized for embedding communication symbols and symbol detection is accomplished by a complicated minimization problem in terms of the permutation matrix.

\section{Regularized Selection based signaling Strategy for DFRC Systems}
\label{sec:regularized}

The two aforementioned signaling strategies implement an unrestricted antenna selection, that is an arbitrary $K$-antenna sparse array might be selected for waveform transmitting according to the embedded symbols. As there are only $K$ RF front-ends installed in the platform, antenna selection network is required to be capable of connecting an arbitrary subset of $K$ antennas with front-ends. This may put a high pressure on the hardware realization especially when the selected antennas locate far from the front-ends. In order to preserve original radar functions, the MIMO radar receiver is assumed to know the association of the orthogonal waveforms to the transmit antennas for the hybrid selection and permutation scheme. The complete transparency between the two functions may cause practical implementation issues as well. To counteract these implementation issues, we propose a regularized selection based signaling strategy to embed communication symbols into the transmit array configuration in the following. 

\begin{figure}[!ht]
  \centering
    \includegraphics[trim = {6cm 7cm 6cm 3.5cm}, scale=0.4]{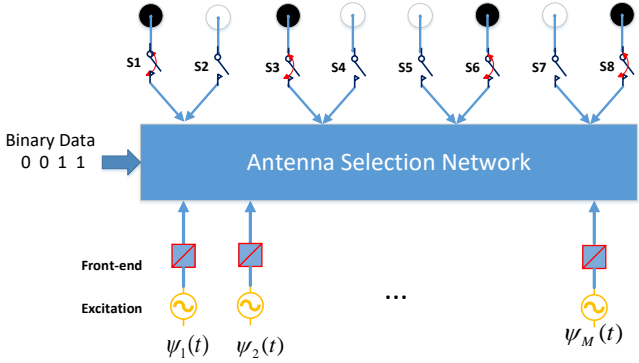}
    \caption{Illustration of regularized antenna selection based modulation signaling scheme.}
\label{fig_2}
\end{figure}

The concept of the proposed signaling scheme is shown in Fig. \ref{fig_2}. There are $M=2K$ uniformly spaced transmit antennas with an inter-element spacing of $d$. The $M$ antennas are divided into $K$ subgroups with each subgroup consisting of two adjacent antennas. Each subgroup represents one-bit symbol, where the symbol ``0'' implies the first antenna selected and the second antenna discarded, and vice versa for the symbol ``1''. The restriction of only one selected antenna for each subgroup can guarantee a constant number of $K$ transmit antennas. Let $\Psi_k(t), m=1,\ldots, K$ be $K$ orthogonal waveforms corresponding to the $K$ subgroups of antennas. The antenna selection matrix in Eq. (\ref{eq:expre_aselected}) becomes  $\bd P(\tau) \in \{0,1\}^{K \times 2K}$, which is a rectangular diagonal selection matrix with a K-bit message matrix $\bd E \in \{0,1\}^{2 \times K}$ populating along the diagonal. Each row $\bd e_k, k=1,\ldots, K$ of the message matrix $\bd E$ is defined as follows:
\begin{equation}
\bd e_k = \begin{cases}
[1, 0] \; \t{if the kth message bit is }b_k = 0, \\
[0, 1] \; \t{if the kth message bit is }b_k = 1. 
\end{cases}
\end{equation}
In order to decouple the dependency of communication performance on the arrival angle $\theta_c$, a set of phase rotations can be pre-multiplied with orthogonal waveforms before transmitting. As proved in section \ref{subsec:ambiguity}, the phase-rotated waveforms are capable of preserving the orthogonality, and do not affect the normal radar operations. To approach the performance of BPSK scheme, the additional phase rotations $\varphi_k, k=1,\ldots,K$ are calculated as,
\begin{equation}
\varphi_k = \begin{cases}
-(2k-2)k_0d\sin\theta_c & \t{if}\; b_k = 0,\\
\pi-(2k-1)k_0d\sin\theta_c  & \t{if}\; b_k = 1.
\end{cases}
\end{equation}
Denoting the phase rotation vector as $\bd u = [e^{j\varphi_1}, \ldots, e^{j\varphi_K}]^T$, the received communication signal is
\begin{equation}
x_c(t,\tau) = \alpha_{\rm ch} \bd a^T(\theta_c) \bd P^T(\tau) \t{diag}(\bd u) \mbd{\Psi}(t) + n_c(t,\tau),
\end{equation}
Matched filtering the received data with the $k$th waveform yields,
\begin{eqnarray}
y_{c,k}(\tau) &=& \int_T x_c(t,\tau) \Psi_k(t) dt, \\
&=& \alpha_{\rm ch} e^{j\varphi_k} \left[\bd a_{2k-1}(\theta_c) (1-b_k) + \bd a_{2k}(\theta_c) b_k \right] \nn\\
&&  +  n_{c,k}(\tau), k = 1, \ldots, K,\nn
\end{eqnarray}
Then, each message bit can be deciphered from the phase of the received signal, that is
\begin{eqnarray}
\label{eq:expre_phasers1}
\hat{\phi}_k(\tau) &=& \t{angle}\{y_{c,k}(\tau)\} - \t{angle}\{ \alpha_{\rm ch} \}, \\
&\approx & \begin{cases}
0 & \t{if} \; b_k = 1,\\
\pi & \t{if} \; b_k = 0, \nn
\end{cases}
\end{eqnarray}
where $\t{angle}(\cdot)$ stands for the angle of a complex number. As the number of embedded bits during each radar pulse equals to the number $K$ of selected antennas, the data rate in bps can be expressed as,
\begin{equation}
R = K \times f_{\t{PRF}}, 
\end{equation}
The bit error rate for the proposed regularized selection strategy is the same as that of BPSK, that is
\begin{equation}
\t{BER}_{r} = Q(\rho,1).
\end{equation}
As the correct detection of the $K$-bit communication symbol requires the accurate estimate of each bit, the symbol error rate of the regularized antenna selection scheme can be expressed in terms of the bit error rate,
\begin{equation}
\t{SER}_{r} = 1-(1-Q(\rho,1))^K.
\end{equation}
Note that there are totally $2^K$ symbols in the dictionary for the regularized selection based signaling scheme and no symbol subset selection is further required. The pair of symbols with the minimum distance is obtained by switching one antenna to the other in one subgroup with maintaining others unchanged and those with the maximum distance is obtained by switching on/off the antennas in all $K$ subgroups. It is worth noting that the association between antennas and orthogonal waveforms is fixed during the entire process and the assumption of communication operation transparency is no more necessary to MIMO radar.

\section{Simulations}
\label{sec:simulations}

In our simulations we consider a radar with $M=16$ antennas arranged in a ULA with an inter-element spacing of $0.25$ wavelength. Throughout the simulations, we assume a number of $K=8$ antennas are selected during each PRI to simultaneously embed one communication symbol while performing the radar operation. The radar receiver array is a 10-antenna ULA. Unless otherwise stated, we evaluate the performance of the system by showing the symbol error rate as a function of SNR. 

\subsection{Example 1: Antenna Selection based signaling Scheme}

\begin{figure}
    \centering
        \includegraphics[trim = {6cm 9cm 6cm 8.8cm}, scale=0.65]{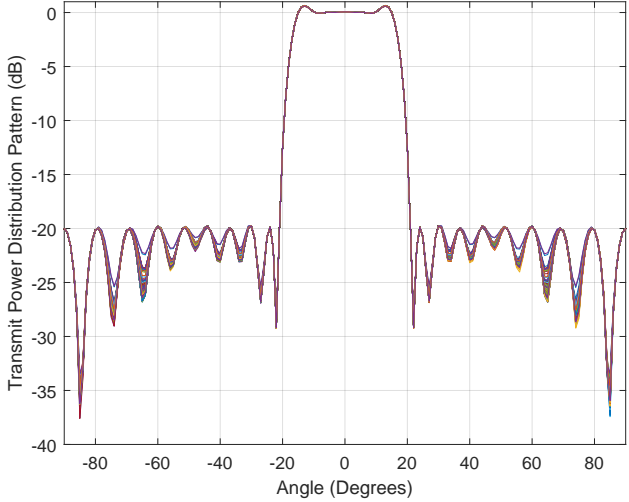} 
        \caption{Overall power patterns of all the 8-antenna sparse arrays in the dictionary $\mathbb{D}_r$.}
        \label{fig_pattern1}
\end{figure}

\begin{figure}
        \centering
        \includegraphics[trim = {6cm 9cm 6cm 8.8cm}, scale=0.65]{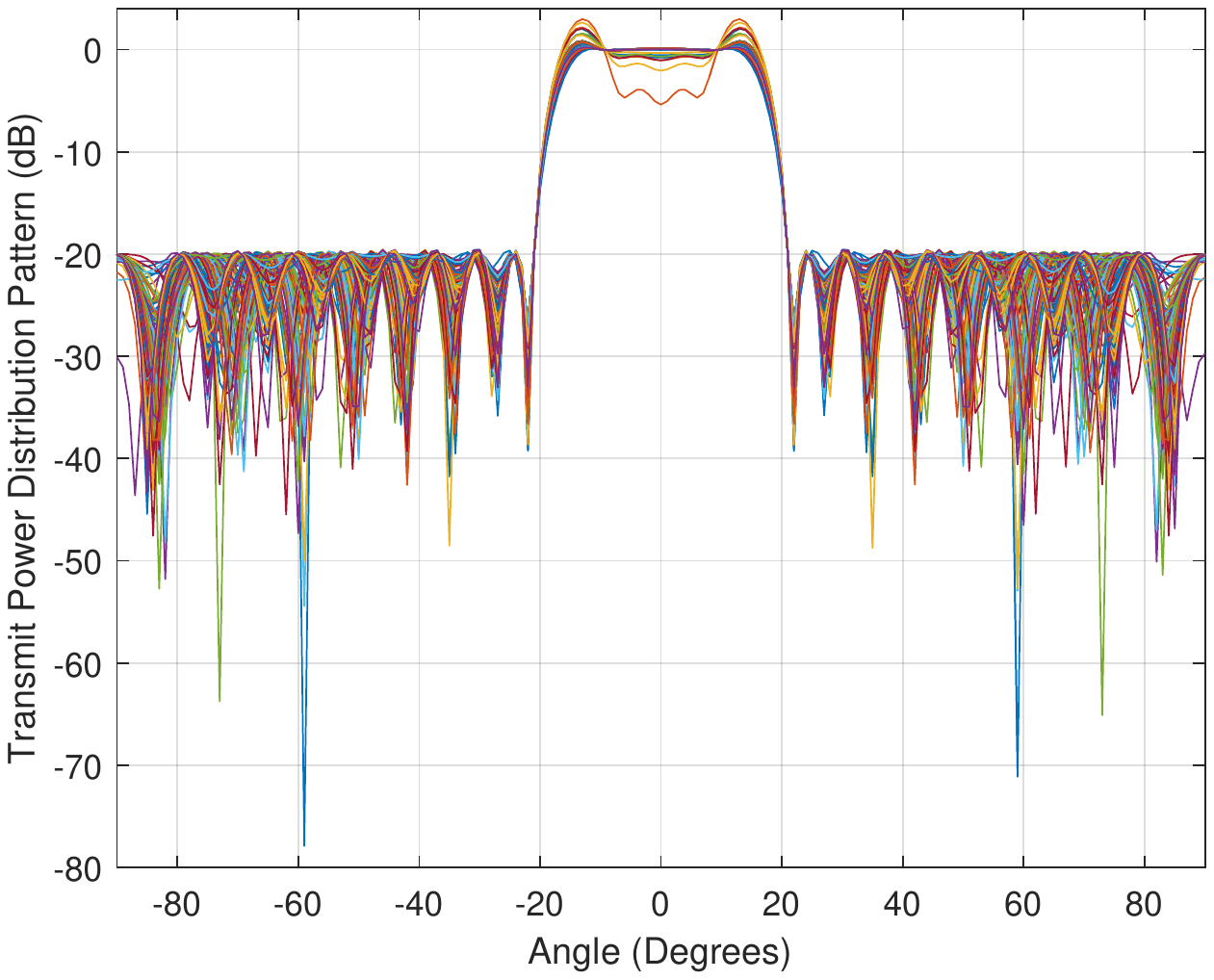} 
        \caption{Overall power patterns of all the 8-antenna sparse arrays in the dictionary $\mathbb{D}_c$.}
        \label{fig_pattern2}
\end{figure}

In the first example, we assume that the main radar operation takes place within the angular sector $\Theta = [-10^{\circ}, 10^{\circ}]$. A single communication receiver is assumed to locate at the direction of $\theta_c=14.4775^{\circ}$. In this case, the total number of unique subarray configurations which can be obtained by antenna selection equals $C_{16}^8=12870$. We embed one communication symbol per PRI. The highest number of bits per symbol is $\lfloor \log_2(C_{16}^8) \rfloor =13$. Here, we consider the cases of 1, 2, 4, and 8 bits per symbol which can be achieved by building four dictionaries of 2, 4, 16, and 256 subarrays, respectively. Symbol subset selection of the 256 configurations, drawn from the total $12870$ available combinations, is performed offline for two scenarios. In the first scenario, the radar operation was given the priority by enforcing the selected sparse arrays to have the smallest peak ripples within the main radar beam. We refer to this set of configurations as $\mathbb{D}_r$. The power patterns of different sparse arrays in the dictionary $\mathbb{D}_r$ are almost the same with a small mainlobe ripple, as shown in Fig. \ref{fig_pattern1}. In the second scenario, we select the sparse arrays such that the Euclidean distance between different symbols in the dictionary is maximized. We refer to this set of configurations as $\mathbb{D}_c$. A peak sidelobe level of $-20$~dB is required in both scenarios.  Unfortunately, the larger Euclidean distance between different symbols comes at the price of having larger difference between the corresponding beampatterns as shown in Fig.~\ref{fig_pattern2}. For small dictionary size, the individual beampatterns have almost the same mainbeam level as the nominal value. However, as the dictionary size increases, some of the individual beampatterns exhibit noticeable deviation from the nominal beampattern within the mainbeam. This may cause some loss in radar performance which is the price paid for having an improved communication detection performance.

\begin{figure}[t]
\centering
\centerline{\epsfig{figure=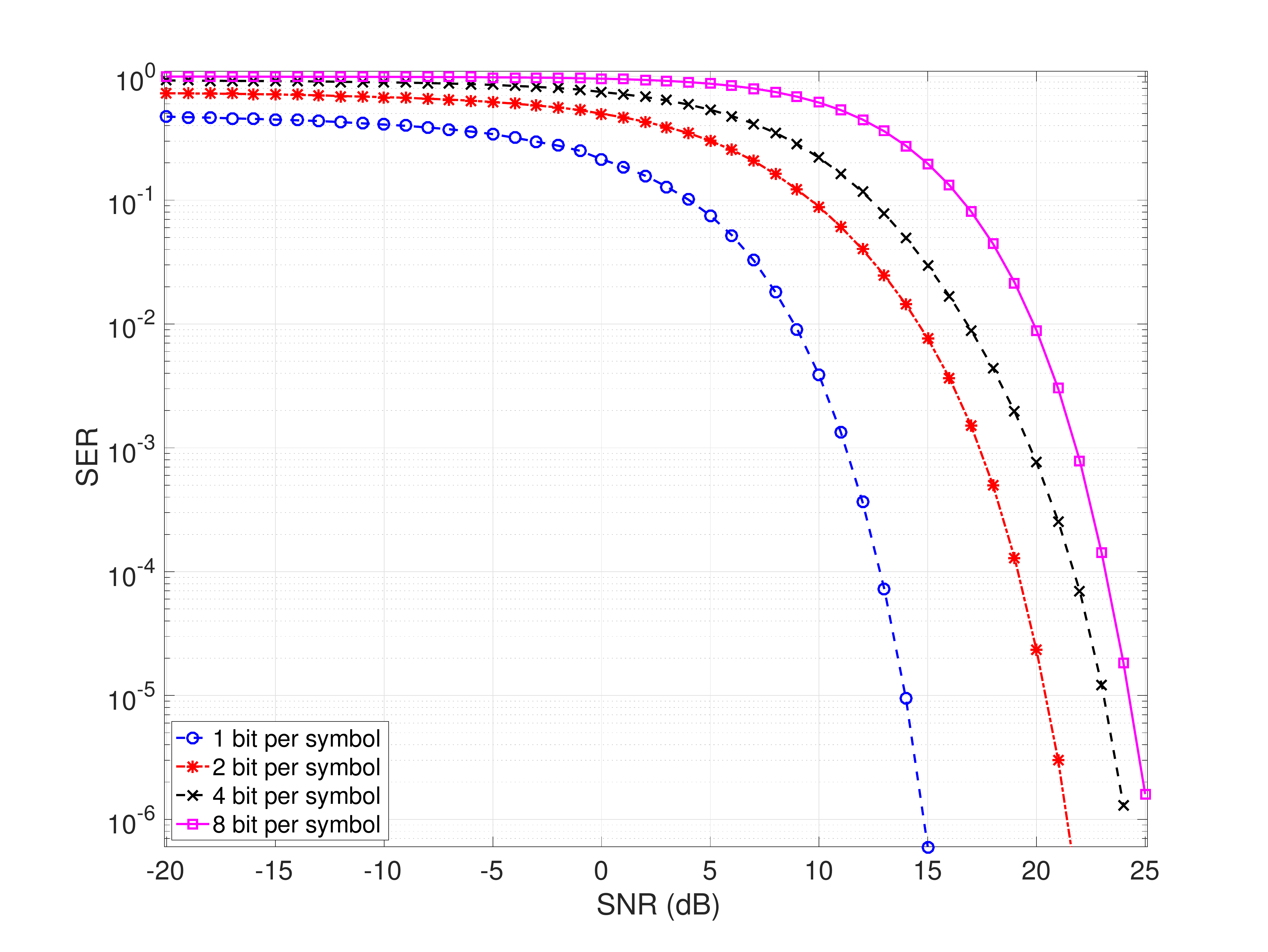,trim = {1cm 1cm 1cm 1.5cm}, scale=0.24}}
\caption{SER versus SNR in the case where the communication receiver is at direction $\theta_c = 14.4775^\circ$; The dictionary $\mathbb{D}_r$ is selected in favor of radar operation.}
\label{fi:SERvsSNR}
\end{figure}
\begin{figure}[!h]
\centering
\centerline{\epsfig{figure=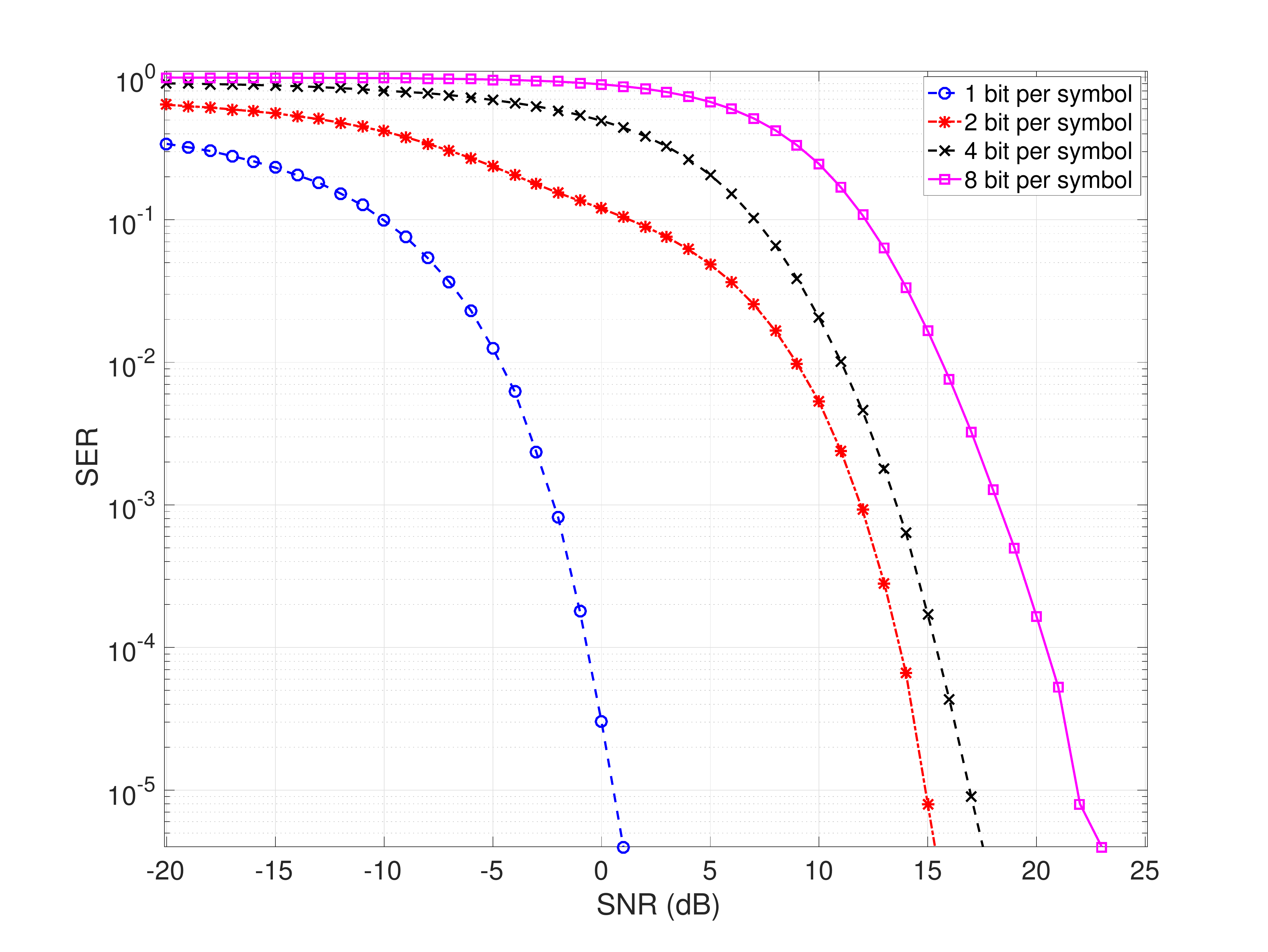,trim = {1cm 1cm 1cm 3cm}, scale=0.24}}
\caption{SER versus SNR in the case where the communication receiver is at direction $\theta_c = 14.4775^\circ$; The dictionary $\mathbb{D}_c$ is selected in favor of communications at the price of increased mainlobe ripples.}
\label{fi:SERvsSNR2}
\end{figure}

To test the communication performance and show the trade-off between the communications and the radar beampattern requirements, a number of $10^7$ symbols are randomly generated. Information embedding is performed using the dictionaries $\mathbb{D}_r$ and $\mathbb{D}_c$, which emphasize the radar requirement and the communication performance, respectively. Figure \ref{fi:SERvsSNR} shows the SER versus SNR for various numbers of bits per symbol using the constellations drawn from $\mathbb{D}_r$. The figure shows that the SER curves exhibit the expected standard behavior of a communication system, with the SER increasing with decreasing SNR and with increasing number of bits per symbol. At high SNR values, a SER smaller than $10^{-5}$ can be achieved for all cases considered. The figure shows that for a fixed SNR the use of a dictionary of smaller size results in lower SER and vice versa. At low SNR values where noise is dominant, the communication receiver detects each symbol in the dictionary with equal probability. For example, the SER is approximately $0.5$ at the SNR of $-20$~dB in the case of 1~bit per symbol, that is the probability of detecting the symbol correctly equals the probability of detecting it erroneously. Figure \ref{fi:SERvsSNR2} shows the SER versus SNR for various numbers of bits per symbol in the scenario where the dictionary $\mathbb{D}_c$ is used. The figure shows that the SER curves exhibit better SER performance as compared to that of the first scenario. This can be attributed to the fact that the dictionary $\mathbb{D}_c$ is designed to enhance the communication performance.

\subsection{Example 2: Hybrid Selection and Permutation based signaling Scheme}

\begin{figure}
    \centering
        \includegraphics[trim = {6cm 9cm 6cm 9.5cm}, scale=0.65]{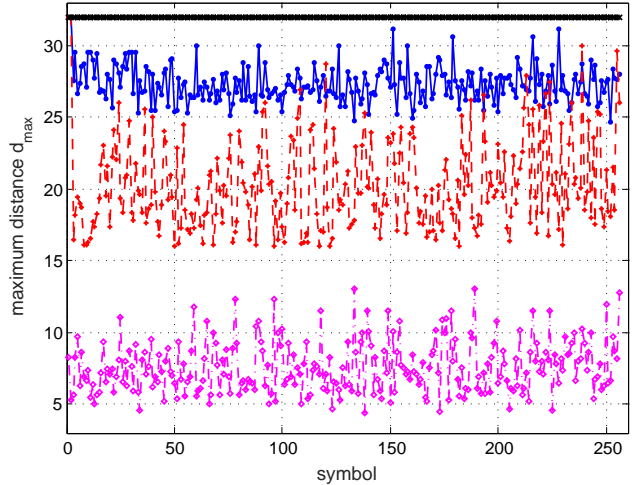} 
        \caption{Maximum distance $d_{\t{max}}^k$ between the $k$th symbol and any other symbol in the dictionaries.}
        \label{fig_distmax}
\end{figure}

\begin{figure}
    \centering
        \includegraphics[trim = {6cm 9cm 6cm 8.5cm}, scale=0.65]{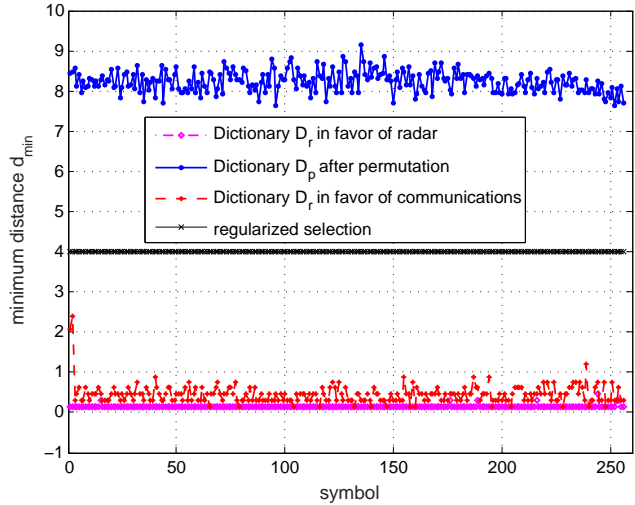} 
        \caption{Minimum distance $d_{\t{min}}^k$ between the $k$th symbol and any other symbol in the dictionaries.}
        \label{fig_distmin}
\end{figure}

We proceed to investigate the hybrid selection and permutation based signaling scheme in this example. For the dictionary $\mathbb{D}_c$ constructed based on the metric of communication performance, we enumerate all potential permutations of each symbol such that the distance between arbitrary two symbols in the dictionary is further maximized. The new dictionary is denoted as $\mathbb{D}_p$. We can calculate the minimum and maximum distances between the $k$th symbol and the remaining 255 symbols as follows,
\begin{equation}
d_{\t{min}}^k  = \begin{cases}
\min\left\lbrace \|\tilde{\bd a}_k - \tilde{\bd a}_i\|^2, i=1, \ldots, k-1, k+1, \ldots, 256 \right\rbrace,  \nn\\
 \t{for} \; \mathbb{D}_c, \mathbb{D}_r; \\
\min\left\lbrace \|\bar{\bd a}_k - \bar{\bd a}_i\|^2, i=1, \ldots, k-1, k+1, \ldots, 256 \right\rbrace, \nn\\
 \t{for} \; \mathbb{D}_p;
\end{cases}
\end{equation}
and
\begin{equation}
d_{\t{max}}^k  = \begin{cases}
\max\left\lbrace \|\tilde{\bd a}_k - \tilde{\bd a}_i\|^2, i=1, \ldots, k-1, k+1, \ldots, 256 \right\rbrace,  \nn\\
\t{for} \; \mathbb{D}_c, \mathbb{D}_r; \\
\max\left\lbrace \|\bar{\bd a}_k - \bar{\bd a}_i\|^2, i=1, \ldots, k-1, k+1, \ldots, 256 \right\rbrace, \nn\\
 \t{for} \; \mathbb{D}_p;
\end{cases}
\end{equation}
The maximum and minimum distances of the constructed three dictionaries $\mathbb{D}_r, \mathbb{D}_c, \mathbb{D}_p$ are plotted in Figs. \ref{fig_distmax} and \ref{fig_distmin} for comparison. Clearly, the minimum distance of the dictionary $\mathbb{D}_p$ after antenna permutation is much larger than those of the two dictionaries, which directly determines the communication accuracy.


\begin{figure}[!h]
\centering
\centerline{\epsfig{figure=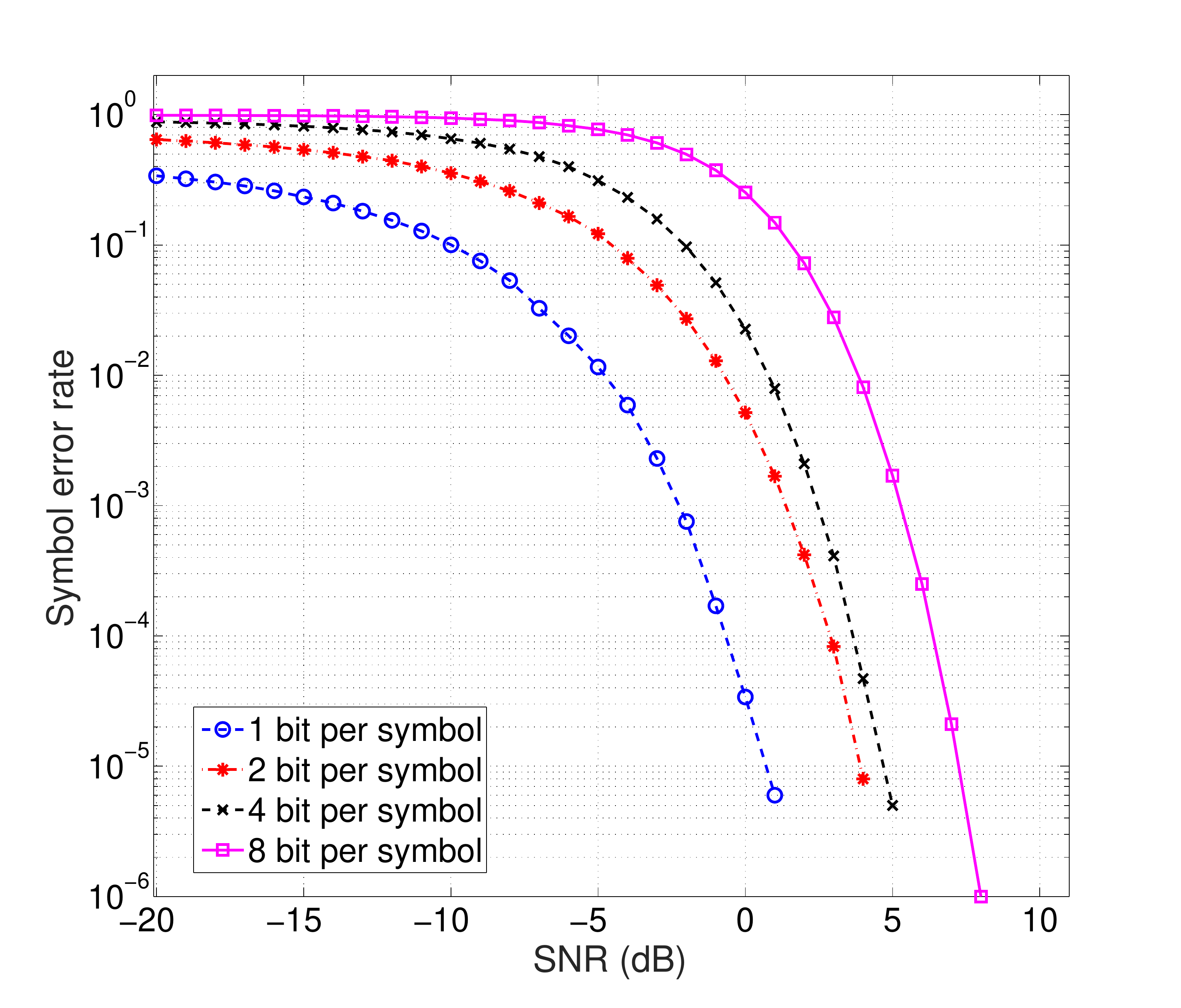,trim = {1cm 1cm 1cm 2cm}, scale=0.3}}
\caption{SER versus SNR in the case where the communication receiver is at direction $\theta_c = 14.4775^\circ$; The dictionary $\mathbb{D}_p$ is obtained by applying permutation to the dictionary $\mathbb{D}_c$ to increase the distance between arbitrary two symbols.}
\label{fi:SERvsSNR3}
\end{figure}

To test the communication performance, a number of $10^7$ symbols are randomly generated. Figure \ref{fi:SERvsSNR3} shows the SER versus SNR for various numbers of bits per symbol. We embed one communication symbol per PRI. The highest number of bits per symbol is $\lfloor \log_2(C_{16}^8 \times 8!) \rfloor =28$. Similar to Example 1, we consider the cases of 1, 2, 4, and 8 bits per symbol, respectively. We can see that the communication performance is significantly improved especially for the case of 8 bits per symbol. 

\subsection{Example 3: Regularized Selection based signaling Scheme}

\begin{figure}
     \centering
     \includegraphics[trim = {6cm 8.8cm 6cm 9cm}, scale=0.7]{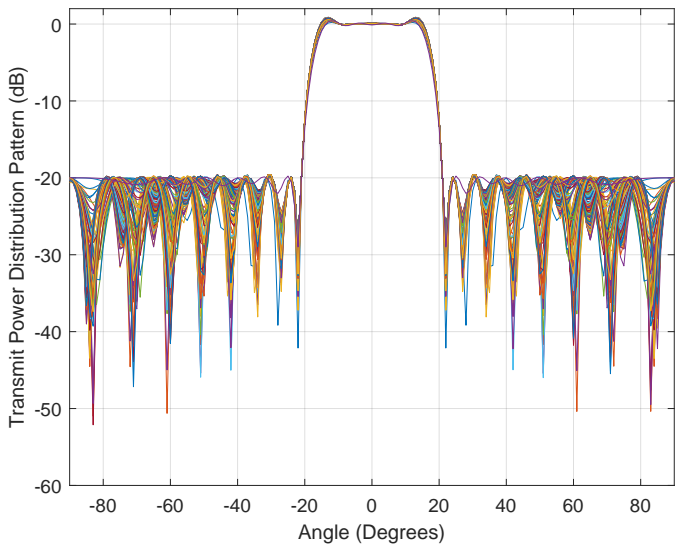} 
     \caption{Power patterns of the 256 different 8-antenna sparse arrays in the regularized selection scheme.}
     \label{fig_pattern3}
\end{figure}

We continue to investigate the regularized selection based signaling scheme. The $16$-antenna ULA is divided into $8$ subgroups and each subgroup consists of two antennas. During each radar pulse, one out of two antennas in each subgroup are switched on according to the communication symbol. There are totally $2^8=256$ symbols and no symbol subset selection is required. The maximum distance is 32, which is obtained by changing the statuses of all $8$ subgroups, as shown in Fig.~\ref{fig_distmax}. The minimum distance is 4, which is achieved by changing the antenna status of one subgroup and maintaining the other subgroups unchanged, as shown in Fig.~\ref{fig_distmin}. The power patterns of the $256$ sparse arrays are depicted in Fig.~\ref{fig_pattern3}, although worse than those of the dictionary $\mathbb{D}_r$ constructed in favor of radar functions, but much better than those of the dictionary $\mathbb{D}_c$ constructed in favor of communication function.

\begin{figure}[!h]
\centering
\centerline{\epsfig{figure=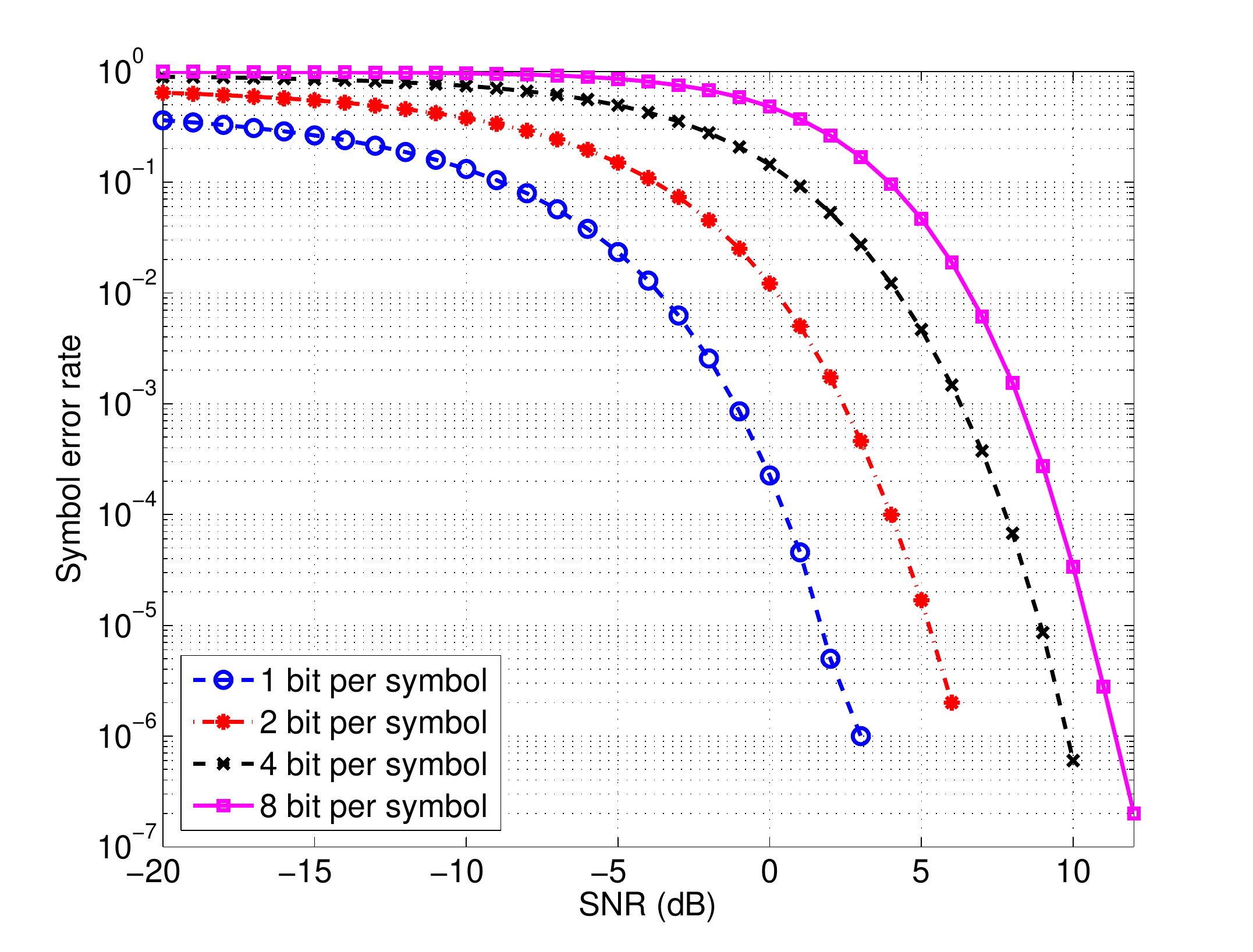,trim = {1cm 1cm 2cm 1cm}, scale=0.45}}
\caption{SER versus SNR for the case where the communication receiver is at direction $\theta_c = 14.4775^\circ$ using the regularized selection scheme.}
\label{fi:SERvsSNR3b}
\end{figure}

To test the communication performance, we consider the cases of 1, 2, 4, and 8 bits per symbol respectively. For the case of 1 bits per symbol, all the 8 subgroups transmit the same bit information. For the case of 2 bits per symbol, the first four subgroups transmit the first bit and the last four subgroups transmit the second bit. For the case of 4 bits per symbol, each two adjacent subgroups transmit one bit information. For the case of 8 bits per symbol, every subgroup transmits one bit. The SER curve versus the SNR is plotted in Fig. \ref{fi:SERvsSNR3b}. Although the communication performance is inferior to that of the hybrid selection strategy, it is much better than those of the antenna-selection scheme with both constellations $\mathbb{D}_r$ and $\mathbb{D}_c$.

\begin{figure}[!h]
\centering
\centerline{\epsfig{figure=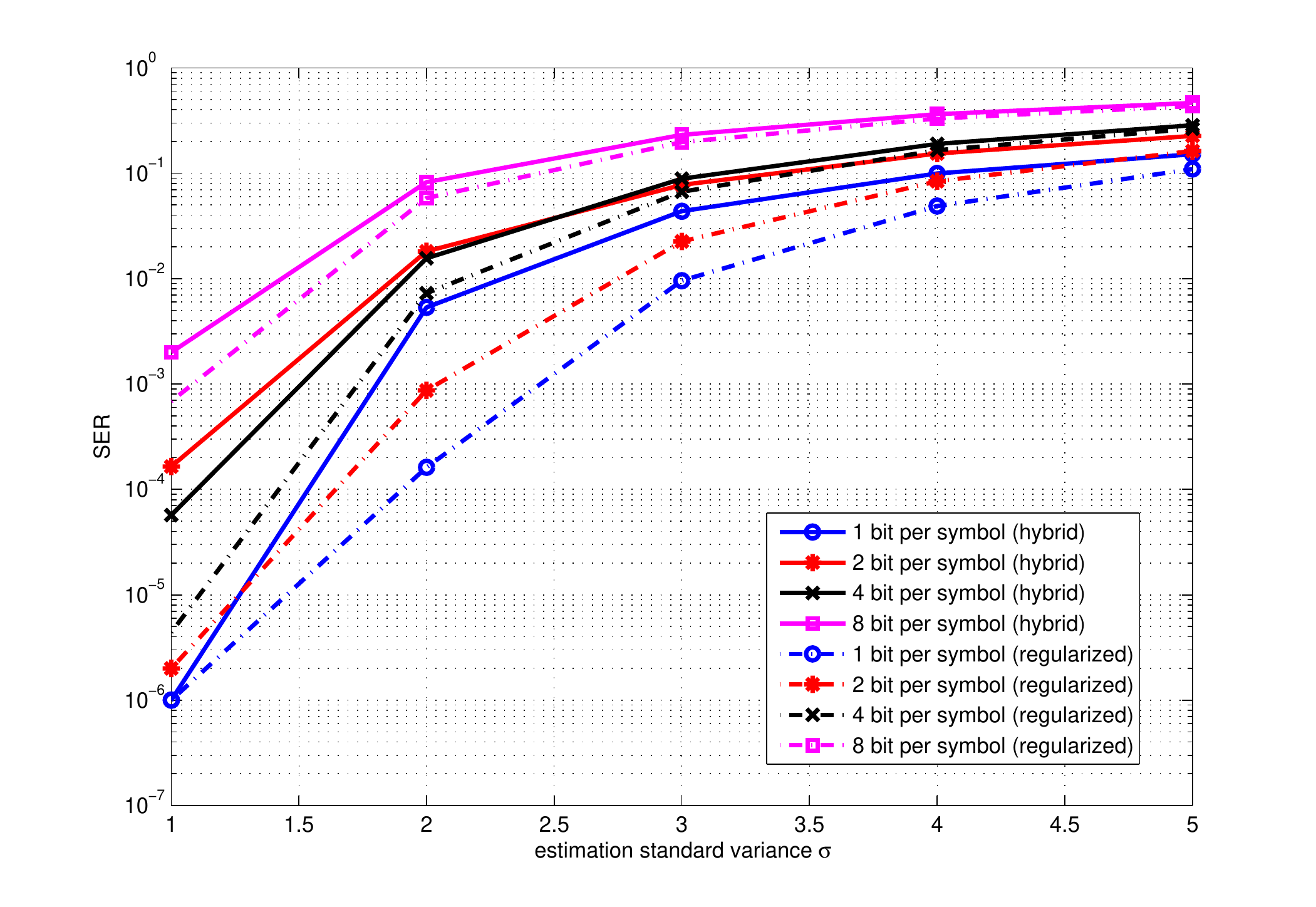,trim = {1cm 1cm 2cm 1cm}, scale=0.45}}
\caption{SER versus the standard variance of communication angle estimation, the communication receiver is assumed at direction $\theta_c = 14.4775^\circ$, the actual angle is normally distributed around $\theta_c$ with variance $\sigma$.}
\label{fi:SERvsVar}
\end{figure}

Finally, we compare the robustness against the estimation error of communication receiver angle between the hybrid scheme and the regularized selection scheme. Assume that the true angle of the communication receiver is normally distributed with mean $\theta_c=14.4775^{\circ}$ and standard variance $\sigma$. The dual-function platform transmits the communication symbol towards the assumed angle $\theta_c=14.4775^{\circ}$ and calculates the phase rotation of each antenna according to that assumed angle. The communication receiver detects the symbol based on the dictionary constructed with the assumed angle. The estimation standard variance $\sigma$ is changing from 1 to 5 in steps of 1 and 500 Monte Carlo simulations are executed for each value. The SER curve versus the standard variance is plotted in Fig.~\ref{fi:SERvsVar} in four cases of 1, 2, 4, and 8 bits per symbol, respectively. We can observe that the regularized selection scheme is more robust against the communication angle estimation error than the hybrid scheme.

\section{Conclusions}
\label{sec:conclusions}

In this paper, we investigated the deployment of sparse arrays by antenna selection for the design of dual functional MIMO radar communications systems. We proposed three new techniques, namely antenna selection, hybrid selection and permutation, and regularized selection based signaling schemes, utilizing transmit array configurations in tandem with waveform diversity for communication information embedding. The strategy of hybrid selection and permutation was able of achieving a megabits high data rate with low symbol error rate. The regularized selection scheme was proposed from the viewpoint of practical implementation and exhibited the best robustness against the estimation error of communication receiver angle. Simulation results validated the successful deployment of sparse arrays in dual functional MIMO radar communications systems for communication performance enhancement without impacting primary radar functions.

\bibliographystyle{ieeetr}
\bibliography{biblio}

\end{document}